\documentclass[11pt]{article}
\usepackage{latexsym}  
\usepackage{amssymb}
\usepackage{graphicx}
\usepackage{amsmath}

\begin{document}

\begin{center}
{\Large\bf Can Massless and Light Yukawaons be Harmless?}

\vspace{5mm}
{\bf Yoshio Koide}

{\it Department of Physics, Osaka University,  
Toyonaka, Osaka 560-0043, Japan} \\
{\it E-mail address: koide@het.phys.sci.osaka-u.ac.jp}

\date{\today}
\end{center}

\vspace{3mm}
\begin{abstract}
In the so-called yukawaon model, where the effective Yukawa 
coupling constants $Y_f^{eff}$ ($f=e,\nu,u,d$) are given
by vacuum expectation values (VEVs) of gauge singlet scalars
(yukawaons) $Y_f$ with $3\times 3$ components, i.e. 
$Y_f^{eff}= y_f \langle Y_f\rangle/\Lambda$, 
massless (and light) scalars appear because a global 
flavor symmetry is assumed. 
In order to demonstrate whether such massless scalars in the 
yukawaon model are harmless or not, yukawaon masses are 
explicitly estimated, as an example, in the charged lepton 
sector.
\end{abstract}

\vspace{3mm}

\noindent{\large\bf 1 \ Introduction}

In the standard model of the quarks and leptons, the 
masse spectra and mixings originate in structures of the 
Yukawa coupling constants $Y_f$ ($f=u,d,\nu,e$), which are
fundamental constants in the theory.
In order to reduce the number of such the fundamental 
constants, usually, we assume flavor symmetries, and
thereby, we discuss relations among the mass spectra
and mixings.
However, even if we assume such symmetries, some 
of such Yukawa coupling constants still remain as the 
fundamental constants in the theory.
On the other hand, 
there is another idea for the origin of the mass spectra 
and mixings which we refer as a yukawaon model
\cite{Yukawaon,e-Yukawaon-PRD09}: 
We regard the Yukawa coupling constants $Y_f$ as ``effective"
coupling constants $Y_f^{eff}$ in an effective theory, 
and we consider that 
$Y_f^{eff}$ originate in vacuum expectation values (VEVs)
of new gauge singlet scalars $Y_f$, i.e.
$$
Y_f^{eff} =\frac{y_f}{\Lambda} \langle Y_f\rangle ,
\eqno(1.1)
$$
where $\Lambda$ is a scale of the effective theory.
We refer the fields $Y_f$ as ``yukawaons".
In the yukawaon model, we can, in principle, calculate the 
VEVs $\langle Y_f\rangle$ from a superpotential 
which is given in the model (although it is not 
yet established at present). 
In the yukawaon model, Higgs scalars are
the same as those in the conventional model, i.e. we
consider only two Higgs scalars $H_u$ and $H_d$ as the
origin of the masses (not as the origin of the
mass spectra).
We also consider that the scale $\Lambda$ is considerably
large (for example, $\Lambda \sim 10^{12}$ GeV).
Therefore, we can inherit successful results from the 
standard model conveniently, and we can introduce 
any flavor symmetries without troubles with 
the gauge SU(2)$_L$ symmetry (a trouble with SU(2)$_L$, 
for example, see Ref.\cite{no-go,no-go-2}).

As a model which describes masses and mixings by patterns
of VEV values of one or more scalars, the Froggatt-Nielsen 
model \cite{Froggatt} is well known: 
The hierarchical structure of the masses is explained 
by a multiplicative structure
$(\langle \phi \rangle/\Lambda)^n$ under a U(1) flavor 
symmetry. 
In contrast to the Froggatt-Nielsen model, in the 
yukawaon model, 
the hierarchical structure of the quark and lepton 
masses is understood from hierarchical eigenvalues
of $\langle Y_f\rangle$, not from the multiplicative
structure $(\langle Y_f\rangle/\Lambda)^n$.
In a supersymmetric (SUSY) yukawaon model, 
the VEVs of yukawaons $\langle Y_f\rangle$ are 
obtained from SUSY vacuum conditions for a superpotential 
$W$, so that the supersymmetry is unbroken at $\mu\sim \Lambda$. 
We will assume that the superpotential $W$ is invariant
under a flavor symmetry.
On the other hand, in order to distinguish a yukawaon $Y_f$ 
from another yukawaons $Y_{f'}$, we will assume a U(1)$_X$
symmetry: 
For example, we assume an O(3) flavor symmetry 
\cite{Koide-O3-PLB08} and we consider that the yukawaons
$Y_f$ are $({\bf 3}\times{\bf 3})_S= {\bf 1}+{\bf 5}$ 
of O(3)$_F$.
Then, the would-be Yukawa interactions are given by
$$
H_{Y}= \sum_{i,j} \frac{y_u}{\Lambda} u^c_i(Y_u)_{ij} {q}_{j} H_u  
+\sum_{i,j}\frac{y_d}{\Lambda} d^c_i(Y_d)_{ij} {q}_{j} H_d 
$$
$$
+\sum_{i,j} \frac{y_\nu}{\Lambda} \ell_i(Y_\nu)_{ij} \nu^c_{j} H_u  
+\sum_{i,j}\frac{y_e}{\Lambda} \ell_i(Y_e)_{ij} e^c_j H_d +h.c. 
+ \sum_{i,j}y_R \nu^c_i (Y_R)_{ij} \nu^c_j ,
\eqno(1.2)
$$ 
where $q$ and $\ell$ are SU(2)$_L$ doublet fields, and
$f^c$ ($f=u,d,e,\nu$) are SU(2)$_L$ singlet fields.
Here, in order to distinguish each $Y_f$ from others, 
we have assigned U(1)$_X$ charges as 
$Q_X(f^c)=-x_f$, $Q_X(Y_f)= +x_f$ and $Q_X(Y_R)=2x_\nu$.
The SU(2)$_L$ doublet fields $q$, $\ell$, $H_u$ and 
$H_d$ have sector charges $Q_X=0$.
As a result of requiring SUSY vacuum conditions,
cross terms among yukawaons $Y_f$
in different sectors $f=u,d,e,\nu$ are allowed 
in the superpotential $W$. 
Therefore, in the yukawaon model, a mass matrix 
$M_f$ in a sector $f$ can be expressed in terms of 
mass matrices in another sectors. 
For example, we can build a model in which the 
Majorana mass matrix $Y_R$ of the right-handed neutrino
is related to an up-quark mass matrix \cite{Koide-O3-PLB08,quark-PLB09}.
This is a significant feature of the yukawaon model.

The original motivation of the yukawaon model was to derive 
a charged lepton mass relation \cite{Koidemass,Koidemass-2,Koidemass-3,K-mass90}.
For example, in the charged lepton sector, we assume
$$
W_e = \lambda_e {\rm Tr}[\Phi_e \Phi_e \Theta_e] + \mu_e {\rm Tr}[Y_e \Theta_e] 
+W_\Phi , 
\eqno(1.7)
$$
where the fields $\Phi_e$ and $\Theta_e$ have the U(1)$_X$
charges $\frac{1}{2} x_e$ and $-x_e$, respectively.
The term $W_\Phi$ in the superpotential (1.7) has been 
introduced in order to fix a VEV spectrum of 
$\langle\Phi\rangle$.
A SUSY vacuum condition $\partial W/\partial \Theta_e=0$
leads to a bilinear mass relation
$$
\langle Y_e\rangle = - \frac{\lambda_e}{\mu_e} 
\langle\Phi_e\rangle \langle\Phi_e\rangle .
\eqno(1.8)
$$
Therefore, the mass spectrum of the charged leptons
are described by \cite{Koide-U3-PLB08,e-spec-PLB09}
$$
K_e \equiv 
\frac{m_e +m_\mu +m_\tau}{(\sqrt{m_e} +\sqrt{m_\mu}
+\sqrt{m_\tau})^2} = \frac{v_1^2+v_2^2+v_3^2}{(
v_1+v_2+v_3)^2} =\frac{{\rm Tr}[\langle\Phi_e\rangle
\langle\Phi_e\rangle]}{{\rm Tr}^2[\langle\Phi_e\rangle]} ,
\eqno(1.9)
$$
and
$$
\kappa_e \equiv \frac{\sqrt{m_e m_\mu m_\tau}}{
(\sqrt{m_e} +\sqrt{m_\mu} +\sqrt{m_\tau} )^3}
=\frac{v_1 v_2 v_3}{(v_1+v_2+v_3)^3} =
 \frac{ \det \langle\Phi_e\rangle}{
{\rm Tr}^3[\langle\Phi_e\rangle]} ,
\eqno(1.10)
$$
where $\langle\Phi_e\rangle={\rm diag}(v_1,v_2, v_3)$.
We refer the field $\Phi_e$ as an ``ur-yukawaon".
Although the mass relation (1.9) with $K_e=2/3$ is 
excellently satisfied with the observed charged lepton 
masses (pole masses), it should be noted that masses 
which we deal with are not ``pole" masses, but ``running" 
masses.
In the yukawaon model, the yukawaons $Y_f$ (and also
ur-yukawaon $\Phi_e$) have their VEVs
$\langle Y_f\rangle$ (and $\langle\Phi_e\rangle$) at 
a high energy scale $\mu=\Lambda$.
In other words, the flavor symmetry O(3) is completely
broken at $\mu=\Lambda$.
Therefore, VEV relations which we obtain from SUSY 
vacuum conditions are valid only at $\mu=\Lambda$.
The effective coupling constants 
$Y_f^{eff}=(y_f/\Lambda)\langle Y_f \rangle$ evolve as 
in the standard model below the scale $\Lambda$.
[The evolution of the relation (1.9) is given, for example,
in Ref.\cite{evol,evol-2}.]
A naive estimate of $\Lambda$ gives 
$m_\nu \sim \langle H_u^0\rangle^2/\Lambda$, i.e.
$$
\Lambda \sim 10^{12}\ {\rm GeV} ,    
\eqno(1.11)
$$
where we have considered that all VEV values (except for
$\langle H_u^0\rangle$ and $\langle H_d^0\rangle$) and 
coefficients $\mu_f$ with a dimension of mass are of 
the order of $\Lambda$. 
Therefore, in the yukawaon model, 
what we should derive is not $K_e(\Lambda)=2/3$,
but $K_e(\Lambda)=(2/3)(1+\varepsilon)$ with 
$\varepsilon \sim 10^{-3}$ \cite{e-spec-PLB09}. 
However, the purpose of the present paper is not
to discuss the value $K_e(\Lambda)$, so that we
do not discuss a reasonable form of $W_\Phi$.

In the present paper, we are interested in massless scalars
which appear when a flavor symmetry is spontaneously broken.
Usually, when a flavor symmetry is spontaneously
broken, the minimum number of the massless scalars is given
by the Goldstone theorem. 
However, in the present yukawaon scenario, there are many 
yukawaons (and ur-yukawaons).
[For example, even in the charged lepton sector, 
we need, at least, three $3\times 3$ scalars, $Y_e$, $\Phi_e$ 
and $\Theta_e$ as seen in the superpotential (1.7).]
It is a great concern to know whose components actually 
become massless, because this is important to discuss its 
physical meaning of the massless (and light) yukawaons.
Regrettably, at present, the whole scenario of 
the yukawaon model is still not completed.
Therefore, in this paper, we investigate only 
massless scalars in the charged lepton sector, because
the sector is a pivotal point of the yukawaon model and 
the results are easily extended to other sectors. 

In order to calculate masses of the yukawaons, we must 
know an explicit expression of the superpotential term
$W_\Phi$.
In this paper, we assume a form 
$$
W_e= \lambda {\rm Tr}[\Phi_e \Phi_e \Theta_e] 
+ \mu {\rm Tr}[Y_e \Theta_e] 
+ \varepsilon_{SB} \left\{ \lambda' {\rm Tr}[\hat{\Phi}_e \Phi_e \Phi_e] 
+ \lambda^{\prime\prime} {\rm Tr}[\hat{\Phi}_e \Phi_e Y_e]
\right\} ,
\eqno(1.12)
$$
where $\hat{\Phi}_e$ is a part of ${\bf 5}$ of O(3) 
$({\bf 3}\times {\bf 3})_S$ and defined by 
$\hat{\Phi}_e=\Phi_e -\frac{1}{3}{\rm Tr}[\Phi_e]$.
Here, we have assigned U(1)$_X$ charges as
$Q_X(Y_e)=x_e$, $Q_X(\Phi_e)=\frac{1}{2}x_e$ and  
$Q_X(\Theta_e)=-x_e$, so that the U(1)$_X$ symmetry is
explicitly broken by the $\varepsilon_{SB}$-terms.
We assume that the value $\varepsilon_{SB}$ is negligibly 
small.
Since we assign the $R$ charges of $\Phi_e$, $Y_e$ and
$\Theta_e$ to $0$, $0$ and $2$, respectively, 
the $\varepsilon_{SB}$-terms also break the $R$ symmetry,
so that the model (1.12) cannot break the supersymmetry
spontaneously \cite{R-sym}.
For the moment, we consider that the supersymmetry is
unbroken, at least, in the Yukawaon sector.

Although the model (1.12) cannot give a realistic charged
lepton mass spectrum at $\mu=\Lambda$ as seen in Appendix, 
however, this is not essential problem for the present 
purpose to estimate massless (and light) yukawaons. 
If we consider an additional term to the superpotential
(1.12) in order to obtain a reasonable mess spectrum
(for example, see Ref.\cite{e-spec-PLB09}), 
the calculation of the yukawaon masses becomes somewhat
complicated, but we can obtain the same conclusions as 
those which we obtain in the next section. 
For simplicity, in the present paper, we will adopt the 
superpotential form (1.12) as a toy model in the yukawaon 
model.

In the next section, we calculate the masses of 
$Y_e$, $\Phi_e$ and $\Theta_e$ based on the yukawaon
model (1.12) in the charged lepton sector.
We will conclude that massless scalars are only three
components $(Y'_{12}, Y'_{23},Y'_{31})$ of a linear 
combination $Y'$ of the scalars 
$Y_e$, $\Phi_e$ and $\Theta_e$.
We will also conclude that the components 
$(Y'_{e 11}, Y'_{e 22}, Y'_{e 33})$ have 
masses of the order of $\varepsilon_{SB} \Lambda$ and 
the remaining components have masses of the order of 
$\Lambda$.
In the section 3, we estimate contributions from yukawaons
in the quark sector, and we check whether the conclusions
only from the lepton sector are modified or not.
Finally, the section 4 is devoted to concluding remarks
(phenomenological effects of the massless yukawaons, 
a value of the scale $\Lambda$, and so on).

\vspace{3mm}

\noindent{\large\bf 2 \ Massless yukawaons}

Our interest is in explicit configurations of massless 
scalars in the present model (1.12),
in which we have three $3\times 3$ scalars $Y_e$, $\Phi_e$
and $\Theta_e$.
Since we consider an unbroken SUSY scenario at present, 
we calculate fermion masses instead of boson masses.
The mass terms is obtained from the superpotential form (1.12) 
as follows:
$$ 
H_{mass}= \lambda {\rm Tr}[\langle \Theta\rangle \Phi\Phi] 
+\lambda {\rm Tr}[\langle \Phi \rangle (\Phi \Theta +\Theta\Phi)] 
+\mu {\rm Tr}[Y\Theta]
$$
$$
+\varepsilon_{SB} \lambda' \left( 3 {\rm Tr}[\langle\Phi \rangle \Phi\Phi] 
-\frac{1}{3} {\rm Tr}[\langle \Phi\rangle] {\rm Tr}[\Phi\Phi] 
-\frac{2}{3} {\rm Tr}[\langle \Phi\rangle\Phi] {\rm Tr}[\Phi] \right)
$$
$$
+\varepsilon_{SB} \lambda^{\prime\prime} \left( 
{\rm Tr}[\langle Y \rangle \Phi\Phi] 
+ {\rm Tr}[\langle\Phi \rangle (\Phi Y + Y \Phi)] 
-\frac{1}{3} {\rm Tr}[\langle \Phi\rangle] {\rm Tr}[\Phi Y]
-\frac{1}{3} {\rm Tr}[\langle \Phi\rangle Y] {\rm Tr}[\Phi]
-\frac{1}{3} {\rm Tr}[\langle Y \rangle\Phi] {\rm Tr}[\Phi] \right)
$$
$$
\equiv 
{\rm Tr}[M_{\Phi\Phi} \Phi\Phi] +{\rm Tr}[M_{\Phi Y} (\Phi Y+Y\Phi)]
+ {\rm Tr}[ M_{\Phi \Theta}(\Phi \Theta+\Theta\Phi)] 
+ {\rm Tr}[M_{\Theta Y} (\Theta Y+Y\Theta)]
$$
$$
+ {\rm Tr}[M_\Phi \Phi +M_Y Y ]{\rm Tr}[\Phi] ,
\eqno(2.1)
$$
where we have dropped index ``$e$" for simplicity, and
$$
M_{\Phi\Phi} = -\varepsilon_{SB} \frac{\lambda^{\prime\prime}
\lambda}{\mu}
\left\{ 2 \langle \Phi \rangle \langle \Phi \rangle 
-\frac{1}{3} {\rm Tr}[\langle \Phi \rangle]\langle \Phi \rangle 
-\xi \left( 3 {\rm Tr}[\langle \Phi \rangle]\langle \Phi \rangle
-\frac{1}{3} {\rm Tr}^2[\langle \Phi \rangle] {\bf 1} \right)
\right\} ,
$$
$$
M_{\Phi Y} = \varepsilon_{SB} \lambda^{\prime\prime} \left( 
\langle \Phi \rangle -\frac{1}{6} {\rm Tr}[\langle \Phi \rangle] 
{\bf 1} \right) , \ \ \ 
M_{\Theta Y}=\frac{1}{2} \mu {\bf 1},
$$
$$
M_\Phi = -\frac{2}{3} \varepsilon_{SB} \lambda' \langle \Phi \rangle
+\frac{1}{3} \varepsilon_{SB}\frac{\lambda^{\prime\prime} \lambda}{\mu}
\langle \Phi \rangle\langle \Phi \rangle , \ \ \ 
M_Y = -\frac{1}{3} \varepsilon_{SB} \lambda^{\prime\prime}
\langle \Phi \rangle .
\eqno(2.2)
$$
We calculate the mass matrix on the basis in which
$\langle \Phi \rangle $ is diagonal, i.e.
$\langle \Phi \rangle={\rm diag}(v_1, v_2, v_3)$.
 
When we define $\Psi^T = (Y, \Phi, \Theta)$, we obtain a 
mass matrix for the components $\Psi_{ij}$ ($i \neq j$) as follows:
$$
H_{mass}=\frac{1}{2} \sum_{i\neq j} \Psi_{ij}^T M_{(ij)} \Psi_{ij} ,
\eqno(2.3)
$$
$$
M_{(ij)} = \left(
\begin{array}{ccc}
0 & M_{\Phi Y}^{(ij)} & M_{\Theta Y}^{(ij)} \\
M_{\Phi Y}^{(ij)} & M_{\Phi \Phi}^{(ij)} &
M_{\Phi \Theta}^{(ij)} \\
M_{\Theta Y}^{(ij)} & M_{\Phi \Theta}^{(ij)} & 0
\end{array} \right) ,
\eqno(2.4)
$$
where
$$
M_{\Phi \Phi}^{(ij)} = (M_{\Phi \Phi})_{ii} +(M_{\Phi \Phi})_{jj}
=-\varepsilon_{SB} \frac{\lambda^{\prime\prime}\lambda}{\mu}
\left\{ 2(v_i^2 +v_j^2) \right.
$$
$$ \left.
-\frac{1}{3}(v_i+v_j)\sum_k v_k
-\xi \left( 3 (v_i+v_j)- \frac{2}{3} {\rm Tr}[\langle \Phi\rangle]\right)
{\rm Tr}[\langle \Phi\rangle] \right\} ,
$$
$$
M_{\Phi Y}^{(ij)} = 
(M_{\Phi Y})_{ii} +(M_{\Phi Y})_{jj} =
\varepsilon_{SB} \lambda^{\prime\prime}
\left( v_i +v_j -\frac{1}{3} {\rm Tr}[\langle \Phi\rangle] \right) ,
$$
$$
M_{\Phi \Theta}^{(ij)} = (M_{\Phi \Theta})_{ii} +(M_{\Phi \Theta})_{jj} 
= \lambda (v_i +v_j) ,
$$
$$
M_{\Theta Y}^{(ij)} =(M_{\Theta Y})_{ii} +(M_{\Theta Y})_{jj} =\mu .
\eqno(2.5)
$$
Since there is no mixing term between $\Psi_{ij}$ and $\Psi_{kk}$,
the mass matrix (2.4) is substantially  a $3\times 3$ matrix.
In order to see whether there is a massless state or not, we
calculate $\det M_{(ij)}$:
$$
\det M_{(ij)} = 2 M_{\Phi Y}^{(ij)} M_{\Phi \Theta}^{(ij)}
M_{\Theta Y}^{(ij)} - M_{\Phi\Phi}^{(ij)} (M_{\Theta Y}^{(ij)})^2 
$$
$$
= \varepsilon_{SB} \lambda^{\prime\prime} \lambda\, \mu \left\{
2(v_i+v_j)^2 +2(v_i^2+v_j^2) - (v_i+v_j){\rm Tr}[\langle \Phi\rangle] 
- \xi \left( 3(v_i+v_j) -\frac{2}{3}{\rm Tr}[\langle \Phi\rangle] \right)
{\rm Tr}[\langle \Phi\rangle] \right\} .
\eqno(2.6)
$$
In this model, we must take $\xi=1$ (see Eq.(A.13) in 
Appendix), so that we obtain
$$
\det M_{(ij)} = \frac{2}{3}  \varepsilon_{SB} \lambda^{\prime\prime} 
\lambda \, \mu \left\{ v_i^2+v_j^2+v_k^2 -4 (v_i v_j + v_j v_k +
v_k v_i) \right\} ,
\eqno(2.7)
$$
where we have used 
$[\langle \Phi\rangle] = \sum_k v_k = v_i+v_j+v_k$.
In the present model, as shown in (A.14) in Appendix, 
the VEV values $v_i$ satisfy
the relation $v_1^2+v_2^2+v_3^2=(2/3)(v_1+v_2+v_3)^2$, i.e.
$v_1^2+v_2^2+v_3^2=4(v_1 v_2+v_2 v_3 +v_3 v_1)$, so that 
the value of (2.7) becomes exactly zero.
We can see that there are no massless states more than three
by calculating ${\rm Tr}[M_{(ij)}]$ and ${\rm Tr}[M_{(ij)}^2]$.

More simply, we can demonstrate it by seeing the case of
$\varepsilon_{SB} \rightarrow 0$. 
[The case $\varepsilon_{SB} \rightarrow 0$ does not mean 
$\varepsilon_{SB} =0$.
The case $\varepsilon_{SB} =0$ means a trivial case with 
$W_\Phi =0$ in Eq.(1.3), so that 
$\langle \Theta _e\rangle = \langle Y_e\rangle =\langle \Phi_e\rangle 
=0$ or $\langle \Theta _e\rangle =0$ and 
$\langle Y_e\rangle =-(\lambda_e/\mu_e) \langle \Phi_e\rangle
\langle \Phi_e\rangle \neq 0$.
Here, the case $\varepsilon_{SB} \rightarrow 0$ means a mass 
matrix $M_{(ij)}$ with $\varepsilon_{SB} \simeq 0$.
]
In this case, the mass matrix $M_{(ij)}$ is given by
$$
M_{(ij)} = \left(
\begin{array}{ccc}
0 & 0 & a \\
0 & 0 & b \\
a & b & 0 
\end{array} \right) ,
\eqno(2.8)
$$
where
$$
a = \mu , \ \ \ b=\lambda (v_i+v_j) .
\eqno(2.9)
$$
The eigenvalues and mixing matrix are given by
$$
m(Y') = 0 , \ \ \ m(\Phi')= -\sqrt{a^2+b^2} , \ \ \ 
m(\Theta ') = \sqrt{a^2+b^2},
\eqno(2.10)
$$
$$
U=\left( 
\begin{array}{ccc}
c & \frac{1}{\sqrt2} s & \frac{1}{\sqrt2} s \\
-s & \frac{1}{\sqrt2} c & \frac{1}{\sqrt2} c \\
0 & -\frac{1}{\sqrt2}  & \frac{1}{\sqrt2}  
\end{array} \right) ,
\eqno(2.11)
$$
respectively, where
$$
s = \frac{a}{\sqrt{a^2+b^2}} , \ \ \ c = \frac{b}{\sqrt{a^2+b^2}} ,
\eqno(2.12)
$$
$$
\left( \begin{array}{c} Y \\ \Phi \\ \Theta  \end{array} \right)
= U \left( \begin{array}{c} Y' \\ \Phi' \\ \Theta ' \end{array} \right) ,
\eqno(2.13)
$$
and $(Y', \Phi', \Theta ')$ are mass-eigenstates.
As seen in Eqs.(2.10) and (2.11), the massless states are
only three, i.e. $(Y'_{12}, Y'_{23},Y'_{31})$, and they 
couple to the charged lepton sector as
$$
\frac{y_e}{\Lambda} \frac{b}{\sqrt{a^2+b^2}} 
\sum_{i\neq j} \ell_i Y'_{ij} e_j^c H_d .
\eqno(2.14)
$$
The situation is almost unchanged for the case $\varepsilon_{SB} \neq 0$
(but $\varepsilon_{SB} \simeq 0$).
Since we consider $\Lambda \sim 10^{12}$ GeV (for details, see
the final section 4), the effective coupling constant of 
$ e_i Y'_{ij} e_j^c$ is of the order of
$\langle H_d\rangle/\Lambda \sim 10^{-13}$ (for $\tan\beta \sim 10$).
The phenomenological meaning will be discussed in the final 
section 4.

Next, we calculate a mass matrix for the components
$\Psi_{ii}$.
Since there are mixing terms between $\Psi_{ii}$ and $\Psi_{jj}$,
i.e. ${\rm Tr}[M_\Phi \Phi + M_Y Y]{\rm Tr}[\Phi]$, the mass matrix is
described by $9\times 9$ matrix for $\Psi_{ii}= (Y_{11},
Y_{22}, Y_{33}, \Phi_{11}, \Phi_{22},\Phi_{33}, \Theta _{11},\Theta _{22},
\Theta _{33})$:
$$
M_{(ii)} = \left(
\begin{array}{ccc}
0 & A & C \\
A & B & D \\
C & D & 0
\end{array} \right) , 
\eqno(2.15)
$$
where $C$ and $D$ take diagonal forms
$$
C=\frac{1}{2} \mu {\bf 1}, \ \ \ 
D= \lambda \, {\rm diag}(v_1, v_2, v_3),
\eqno(2.16)
$$
while $A$ and $B$ have off-diagonal elements as
$$
A_{ii,ii}= \varepsilon_{SB} \lambda^{\prime\prime} 
\left( v_i -\frac{1}{6} {\rm Tr}[\langle \Phi\rangle] \right) , \ \ \ 
A_{ii,jj} = -\frac{1}{6} \varepsilon_{SB} \lambda^{\prime\prime} 
(v_i+v_j) ,
\eqno(2.17)
$$
$$
B_{ii,ii}= -\varepsilon_{SB} \frac{\lambda^{\prime\prime}\lambda}{\mu}
\left( 2v_i^2 -\frac{1}{3} v_i {\rm Tr}[\langle \Phi\rangle] 
- 3\xi v_i \sum_k v_k
+\frac{1}{3}\xi {\rm Tr}[\langle \Phi\rangle]^2 \right) ,
$$
$$
B_{ii,jj} = \frac{1}{6} \varepsilon_{SB} 
\frac{\lambda^{\prime\prime}\lambda}{\mu} \left(
v_i^2 +v_j^2 - 2 \xi (v_i+v_j){\rm Tr}[\langle \Phi\rangle] \right).
\eqno(2.18)
$$
In order to see whether there are massless states or not,
we calculate $\det M_{(ii)}$:
$$
\det M_{(ii)} = \frac{1}{2(24)^3} (\varepsilon_{SB} 
\lambda^{\prime\prime}\lambda\, \mu {\rm Tr}^2[\langle \Phi\rangle])^3 
\left( -171K^3 -1053 K^2 +2283 K -947 \right.
$$
$$ \left.
+93092 \kappa^2
-17526 \kappa + 32846 \kappa K \right) ,
\eqno(2.19)
$$
for $\xi=1$.
For the value $K=2/3$, we obtain
$$
\det M_{(ii)} = \frac{1}{6(24)^3} (\varepsilon_{SB} 
\lambda^{\prime\prime}\lambda\, \mu {\rm Tr}^2[\Phi])^3 \left(
169+ 1114 \kappa+ 372368 \kappa^2 \right) ,
\eqno(2.20)
$$
so that the value (2.20) can never become zero
either for a positive $\kappa$ [see (A.17) in Appendix] 
or for the negative value $\kappa=-1/54$ given in Eq.(A.15).
Therefore, we cannot have massless scalars for
the components $\Psi_{ii}$.

However, in the limit of $\varepsilon_{SB} \rightarrow 0$,
the mass matrix $M_{(ii)}$ becomes the same type with
(2.8), so that $Y'_{(ii)}$ become massless.
Therefore, by taking the result (2.20) into consideration,
we can conclude that  the masses of $Y'_{ii}$ are given by
$$
m(Y'_{ii}) \sim \varepsilon_{SB} \lambda^{\prime\prime}
\lambda\, \mu \frac{ {\rm Tr}^2[\langle\Phi\rangle]}{\Lambda^2}
 \sim \varepsilon_{SB} \Lambda .
\eqno(2.21)
$$
If we consider $\varepsilon_{SB} \sim 10^{-12}$, it is possible
that the masses of $Y'_{ii}$ appear in a TeV region.
However, at present, we do not fix the order of 
$\varepsilon_{SB}$.

In conclusion, we have strictly calculated masses of 
yukawaons in the charged lepton sector based on a 
superpotential form (1.12).
We have found that, of the fields
$Y_e$, $\Phi_e$ and $\Theta _e$, the massless components
are only three $(Y'_{e 12}, Y'_{e 23}, Y'_{e 31})$.
The components $(Y'_{e 11}, Y'_{e 22}, Y'_{e 33})$ have 
masses of the order of $\varepsilon_{SB} \Lambda$ and 
the remaining components have masses of the order of 
$\Lambda$.


\vspace{3mm}

\noindent{\large\bf 3 \ Contribution from quark sector}

As far as we see the results in the charged
lepton sector, there is no mixing between different components 
$\Psi_{ij}$ and $\Psi_{kl}$ (but there are mixings 
between $\Psi_{ii}$ and $\Psi_{jj}$). 
However, this is not true when we consider
whole yukawaons in the all sectors.
For example, in order to give a nearly tribimaximal mixing
\cite{tribi,tribi-2,tribi-3,tribi-4,tribi-5,tribi-6},
the author \cite{Koide-O3-PLB08} has proposed superpotential 
terms
$$
W_R =\mu_R {\rm Tr}[Y_R \Theta_R] + \lambda _R {\rm Tr}[
(\Phi_u Y_e + Y_e \Phi_u)\Theta _R].
\eqno(3.1)
$$
where $\langle \Phi_u \rangle$ is related to up-quark mass 
matrix $M_u$ via a relation 
$$
\langle Y_u\rangle = - \frac{\lambda_u}{\mu_u} 
\langle\Phi_u\rangle \langle\Phi_u\rangle .
\eqno(3.2)
$$
from superpotential terms in up-quark sector
$$
W_u = \lambda_u {\rm Tr}[\Phi_u\Phi_u \Theta _u]
+\mu_u {\rm Tr}[Y_u \Theta _u] ,
\eqno(3.3)
$$
similar to Eq.(1.7) (neglecting $W_{\Phi_u}$).
Then, the terms (3.1) contribute to the yukawaon mass terms as
$$
\lambda_R {\rm Tr}[\langle \Theta _R \rangle (Y_e \Phi_u +\Phi_u Y_e)]
+ \lambda _R {\rm Tr}[\langle\Phi_u\rangle (Y_e \Theta _R +\Theta _R Y_e) ].
\eqno(3.4)
$$
In this case, $\langle \Theta _R \rangle$ is zero \cite{Koide-O3-PLB08},
but $\langle\Phi_u\rangle$ is not zero, and besides,
$\langle\Phi_u\rangle$ is not diagonal on the diagonal 
basis of $\langle\Phi_e\rangle$.
The existence of the terms (3.1) causes $Y'_{ij}$-$Y'_{jk}$ 
mixings among the massless yukawaon $Y'_e$ described in
Eq.(2.10).
In other words, the flavor-changing neutral currents 
(FCNC) via $Y'_{ij}$ can appear in principle.
From superpotential terms in the up-quark sector
From the superpotential (3.3), we also obtain mass terms
$$
\lambda_u {\rm Tr}[\langle \Theta _u\rangle \Phi_u\Phi_u
+\langle \Phi_u\rangle (\Phi_u \Theta _u +\Theta _u \Phi_u)]
+\mu_u {\rm Tr}[Y_u \Theta _u].
\eqno(3.5)
$$
As a result, we obtain the following mass matrix for
$\Psi^{(u)} =(Y_u, \Phi_u, \Theta _u, Y_e, Y_R, \Theta _R)$
$$
M^{(u)}= \left(
\begin{array}{cccccc}
0 & 0 & d_u & 0 & 0 & 0 \\
0 & 0 & a & 0 & 0 & 0 \\
d_u & a & 0 & 0 & 0 & 0 \\
0  & 0 & 0 & 0 & 0 & c \\
0  & 0 & 0 & 0 & 0 & d_R \\
0  & b & 0 & c & d_R & 0 
\end{array} \right) ,
\eqno(3.6)
$$
where
$$
a=\lambda_u \langle \Phi_u\rangle , \ \ 
b= \lambda_R \langle Y_e \rangle , \ \ 
c= \lambda_R \langle \Phi_u\rangle , \ \ 
d_u = \frac{1}{2} \mu_u {\bf 1} , \ \ 
d_R = \frac{1}{2} \mu_R {\bf 1} ,
\eqno(3.7)
$$
and we have put $\langle \Theta _u\rangle =\langle \Theta _R\rangle=0$.
Although the matrix $b$ is not diagonal on the diagonal 
basis of $\langle \Phi_u\rangle$, in order to obtain
a rough sketch of the mixing, we regard the matrix (3.7)
as a $6\times 6$ matrix.
Then, we obtain two massless states $Y'_u$ and $Y'_e$,
which have the following components
$$
Y'_u = \frac{a d_R}{N_u} Y_u
- \frac{b d_R}{N_u} \Phi_u
+ \frac{b d_u}{N_u} Y_R,
\eqno(3.8)
$$
$$
Y'_e = \frac{a c}{N_e} Y_u
- \frac{c d_u}{N_e} \Phi_u
+ \frac{b d_u}{N_e} Y_e,
\eqno(3.9)
$$
respectively, where $N_u =\sqrt{(ad_R)^2+(bd_R)^2+(bd_u)^2}$
and $N_e=\sqrt{(ac)^2+(cd_u)^2+(bd_u)^2} $.
The state $Y'_e$ in (3.9) is not identical with $Y'_e$ 
given in (2.13).
The former is defined as a massless state in the mass
matrix for 
$\Psi^{(u)} =(Y_u, \Phi_u, \Theta _u, Y_e, Y_R, \Theta _R)$,
and the latter is defined as a massless state in the
mass matrix for $\Psi =(Y_e, \Phi_e, \Theta _e)$.
Of course, the exact mass-eigenstates should be calculated
by diagonalizing a mass matrix for whole yukawaons 
simultaneously.
Nevertheless, from the results (3.8) and (3.9), 
we can deduce the following overviews:
(i) The massless particle $Y'_u$ contains
a component $Y_u$, but it does not contain $Y_e$, while
the massless particle $Y'_e$ contains both components
$Y_u$ and $Y_e$.
(ii) We consider that $Y'_u$ has a mass of the order
$\varepsilon_{SB} \Lambda$ due to terms which we have neglected
in (3.2) (which is corresponding to $W_{\Phi_u}$),
but $Y'_e$ will be still massless, so that only three
massless components $(Y'_{12}, Y'_{23}, Y'_{31})$
can couple to all quark and lepton sectors.
Here, we have dropped the index ``$e$" from
the latter particles $Y'_e$ because the particles
$(Y'_e)_{ij}$ couple not only to the charged lepton 
sector but also to whole quark and lepton sectors
(except for the $\nu^c_i \nu^c_j$ sector). 
(iii) Since the VEV matrices $\langle Y_u\rangle $ and 
$\langle Y_e\rangle $ cannot
simultaneously be diagonalized, the indexes $(1,2,3)$
in $(Y'_{12}, Y'_{23}, Y'_{31})$ do not mean 
either $(1,2,3)=(e,\mu,\tau)$ or $(1,2,3)=(u,c,t)$.
Therefore,  FCNC effects can appear via exchanges 
of $Y'$ in principle.

\vspace{3mm}

\noindent{\large\bf 4 \ Concluding remarks}

In conclusion, we have estimate yukawaon masses from
the superpotential terms (1.12) and (3.1) explicitly,
and we have obtained three massless yukawaons 
$(Y'_{12}, Y'_{23}, Y'_{31})$ and three light yukawaons 
$(Y'_{11}, Y'_{22}, Y'_{33})$.
The superpotential term is only a toy model, the 
results seem to be general ones. 
Finally, we would like to check whether those 
particles are harmless or not in phenomenology (except for
cosmological problems). 

For example, massless yukawaon $Y'_{12}$ couple to
charged lepton sector, so that it causes a muon decay 
$\mu \rightarrow e +Y'_{12}$.
However, it is invisibly small compared with 
the weak decay $\mu \rightarrow e +\bar{\nu} +\nu$, 
because the effective coupling constant of the term
$\ell Y_e e^c)$ is given by 
$$
g\equiv y_e \frac{\langle H_d\rangle}{\Lambda} \sim 10^{-11} ,
\eqno(4.1)
$$ 
where we have taken $v_d \equiv \langle H_d \rangle \sim 10^1$
GeV (for $\tan\beta \sim 10$) and $\Lambda \sim 10^{12}$ GeV.
Similarly, the massless yukawaons  $(Y'_{12}, Y'_{23}, Y'_{31})$ 
can couple not only to the charged lepton sector
but also to other quark and lepton sectors.
However, production rates of $Y'_{ij}$ and decay rates 
of the conventional quarks and leptons  into $Y'_{ij}$
are invisibly small because of 
the too small effective coupling constant (4.1).

By the way, can we regard the light yukawaons 
$(Y'_{11}, Y'_{22}, Y'_{33})$ with masses of the order 
of $\varepsilon_{SB} \Lambda$ as a candidate of the dark
matter? 
Since we have interaction terms
$\varepsilon_{SB} \lambda' {\rm Tr}[\hat{\Phi}_e\Phi_e\Phi_e] 
+\varepsilon_{SB} \lambda^{\prime\prime} {\rm Tr}[\hat{\Phi}_e\Phi_e Y_e]$ 
in Eq.(1.12), we can have physical interaction terms 
$Y'_{ij}Y'_{jk} Y'_{ki}$,
so that the light yukawaons $Y'_{ii}$ can decay into
$\tilde{Y}'_{ij} + \tilde{Y}'_{ij}$ with the effective
coupling constant of the order $\varepsilon_{SB}$, where 
$\tilde{Y}'$ denotes a SUSY partner of the scalar $Y'$.
(Although we have interactions $\lambda {\rm Tr}[\Phi_e\Phi_e \Theta _e]$
without the factor $\varepsilon_{SB}$ in Eq.(1.12), the 
field $\Theta _e$ does not contain either the massless yukawaons
or the light yukawaons, the effective coupling constants
of $Y'Y'Y'$ cannot be given by $\lambda$ without the factor 
$\varepsilon_{SB}$.)
Since $Y'_{ii}$ have masses $m_{Y'}$ of the order of
$\varepsilon_{SB} \Lambda$ as seen in Eq.(2.21), the decay widths 
are of the order
of $\varepsilon_{SB}^2 m_{Y'}\sim \varepsilon_{SB}^3 \Lambda$. 
At present, the parameter $\varepsilon_{SB}$ is free.
Therefore, the lifetimes can freely be adjusted. 
However, regrettably, we cannot regard the light yukawaons
as candidates of the cold dark matter:
In order to get a small decay width of the order $10^{-43}$ GeV
which is required from the age of the universe,
we must consider
$$
10^{-45}\ {\rm GeV} \sim \Gamma (Y'_{ii}) \sim 
\frac{(\varepsilon_{SB})^2}{4\pi} m(Y'_{ii})
\sim 10^{-1} (\varepsilon_{SB})^3 \Lambda , 
\eqno(4.2)
$$
which leads to $\varepsilon_{SB} \sim 10^{-18}$ for
$\Lambda \sim 10^{12}$ GeV, so that we get 
$m(Y'_{ii}) \sim 10^{-6}$ GeV.
This value of $m(Y'_{ii}) $ is too small to regard $'_{ii}$
as a candidate of the cold dark matter.

However, here, let us recheck the estimate of $\Lambda$, (1.11).
The result $\Lambda \sim 10^{12}$ GeV is obtained
as follows:
From the relation (1.1) in the charged lepton sector, 
we estimate
$$
\frac{\langle Y_e \rangle}{\Lambda} \sim 
\frac{ 1\ {\rm GeV}}{y_e v_d} \sim \frac{1}{y_e} 10^{-1},
\eqno(4.3)
$$
where we have assumed $\tan\beta\equiv v_u/v_d \sim 10$,
so that $v_d \equiv \langle H_d \rangle \sim 10$ GeV.
If we consider $y_e \sim 10^{-1}$, we can consider 
$\langle Y_e \rangle \sim \Lambda$.
On the other hand, from the seesaw neutrino mass matrix
$M_\nu = m_D M_R^{-1} m_D^T$, where $m_D = (y_\nu/\Lambda)
\langle Y_\nu \rangle v_u$ and $M_R =y_R \langle Y_R\rangle
\sim \Lambda_R$, we estimate
$$
\Lambda_R \sim \frac{( 1\ {\rm GeV})^2}{10^{-10}\ {\rm GeV} }
 \left(\frac{v_u}{v_d}\right)^2
\left(\frac{y_\nu}{y_e}\right)^2 \sim 10^{12} {\rm GeV} ,
\eqno(4.4)
$$
where we have taken $m_{\nu 3} \sim 10^{-10}$ GeV and 
$({y_\nu}/{y_e})^2 \sim 1$.
(In a neutrino model given in Ref.\cite{YK-JIMPA09}, 
the U(1)$_X$ charges
of $e^c$ and $\nu^c$ are the same, so that the yukawaon
$Y_e$ couples not only to the charged lepton sector, but
also to the neutrino sector, and we can consider a model
without $Y_\nu$.)
Note that the scale $\Lambda$ is independent of the scale
$\Lambda_R$ at present.

The relation between $\Lambda$ and $\Lambda_R$ comes from
the relation (3.1):
$$
\Lambda_R \sim \frac{1}{\mu_R} \langle \Phi_u\rangle
\langle Y_e \rangle \sim \frac{\Lambda^2}{\mu_R} .
\eqno(4.5)
$$
The simplest interpretation of (4.5) is to consider
$$
\Lambda_R \sim \Lambda \sim \mu_R \sim 10^{12}\ 
{\rm GeV}.
\eqno(4.6)
$$
Thus, the conclusion (1.11) has been obtained.

However, in a recent version of the yukawaon model 
in the neutrino sector, the author \cite{quark-PLB09} 
has proposed a modified superpotential terms
$$
W_R =\mu_R {\rm Tr}[Y_R \Theta_R] + \frac{\lambda_R}{\Lambda} 
{\rm Tr}[(\Phi_u P_u Y_e + Y_e P_u \Phi_u)\Theta _R],
\eqno(4.7)
$$
where $P_u$ is a new field which appears only in the right-handed
neutrino sector.
If we regard the order of $\langle P_u\rangle$ as  
$\langle P_u\rangle \sim \Lambda_R$, the relation (4.5) is 
replaced with
$$
\Lambda_R \sim \frac{1}{\mu_R \Lambda}\langle \Phi_u\rangle
 \langle Y_e \rangle \langle P_u\rangle
 \sim \frac{\Lambda^2\Lambda_R}{\mu_R \Lambda} .
\eqno(4.8)
$$
Then, if we regarded $\mu_R$ as $\mu_R \sim \Lambda$,
the scale $\Lambda$ can be independent of the 
scale $\Lambda_R$, so that we can take a lower value 
than $\Lambda_R \sim 10^{12}$ GeV.
Of course, a scale of $\Lambda$ with too low value is
dangerous phenomenologically. 
A case with a lower value $\Lambda$ will be investigated
elsewhere.

Thus, we can conclude that the massless and light
yukawaons are harmless in low and high 
energy physics, as far as we consider $\Lambda \sim 10^{12}$
GeV. 
In other words, regrettably, it is hard to find a positive
effect of these yukawaons in low and high energy physics
experiments.
On the other hand, in cosmology, it is likely that the 
existence of such massless yukawaons causes problems.
(Besides, there is a possibility that the massless 
yukawaons acquire unwelcome masses due to radiative 
corrections.)
If such troubles turn out to be serious, we will be 
obliged to consider the flavor symmetry as a gauged one.
     

\vspace{6mm}

\centerline{\large\bf Acknowledgments}

The author would like to thank T.~Yamashita for 
careful checking an earlier calculation of the yukawaon masses 
and pointing an mistake in the calculation. 
The author also would like to thank Y.~Hyakutake,
M.~Tanaka, N.~Uekusa and R.~Takahashi for helpful discussions 
on the massless yukawaons and their phenomenological meaning.
This work is supported by the Grant-in-Aid for
Scientific Research (C), JSPS, (No.21540266).

\vspace{3mm}

\begin{center}
{\large\bf Appendix: VEV structure of the ur-yukawaon} 
\end{center}

In this appendix, we discuss a VEV structure of the
superpotential (1.12), i.e.
$$
W_e= \lambda {\rm Tr}[\Phi_e \Phi_e \Theta_e] 
+ \mu {\rm Tr}[Y_e \Theta_e] 
 + \varepsilon_{SB} \lambda' \left( 
{\rm Tr}[\Phi_e\Phi_e\Phi_e] 
-\frac{1}{3}{\rm Tr}[\Phi_e]{\rm Tr}[\Phi_e\Phi_e] 
\right) 
$$
$$
+ \varepsilon_{SB} \lambda^{\prime\prime} \left( 
{\rm Tr}[\Phi_e\Phi_e Y_e] 
-\frac{1}{3}{\rm Tr}[\Phi_e]{\rm Tr}[\Phi_e Y_e] 
\right)  .
\eqno(A.1)
$$
Here, we have assumed that the U(1)$_X$ symmetry is 
explicitly broken by the order $\varepsilon_{SB}$ which is 
negligibly small.

SUSY vacuum conditions  
$\partial W/\partial Y_e=0$ and 
$\partial W/\partial \Phi_e=0$ lead to 
$$
\Theta_e = -\varepsilon_{SB} \frac{\lambda^{\prime\prime}}{
\mu} \hat{\Phi}_e \Phi_e ,
\eqno(A.2)
$$
and 
$$
\frac{\partial W}{\partial \Phi_e} = 
\lambda (\Phi_e \Theta_e + \Theta_e \Phi_e ) 
+ \varepsilon_{SB} \lambda ' \left( 3 \Phi_e^2 
- \frac{2}{3}{\rm Tr}[\Phi_e] \Phi_e 
-\frac{1}{3}{\rm Tr}[\Phi_e\Phi_e] {\bf 1} \right)
$$
$$
+\varepsilon_{SB} \lambda^{\prime\prime} \left(
\Phi_e Y_e +Y_e \Phi_e -\frac{1}{3} {\rm Tr}[\Phi_e] Y_e
-\frac{1}{3} {\rm Tr}[\Phi_e Y_e] {\bf 1} \right)
= 0 ,
\eqno(A.3)
$$
respectively.
By substituting (A.2) into (A.3), we obtain a cubic 
equation for $\langle\Phi_e \rangle$: 
$$
c_3 \langle\Phi_e\rangle^3 + c_2 \langle\Phi_e\rangle^2 
+c_1 \langle\Phi_e\rangle + c_0 {\bf 1} = 0 .
\eqno(A.4)
$$
where 
$$
c_3 = 4, \ \  c_2= -(1+3\xi){\rm Tr}[\Phi_e] , \ \ 
c_1=\frac{2}{3}\xi {\rm Tr}^2[\Phi_e] , \ \ 
c_0 = -\frac{1}{3} \left( {\rm Tr}[\Phi_e \Phi_e \Phi_e]
-\xi {\rm Tr}[\Phi_e]{\rm Tr}[\Phi_e \Phi_e] \right) ,
\eqno(A.5)
$$
and $\xi$ is defined by
$$
\xi = \frac{\lambda'}{\lambda^{\prime\prime}}
\frac{\mu}{\lambda {\rm Tr}[\Phi_e]} .
\eqno(A.6)
$$

The coefficients $c_a$ ($a=0,1,2,3$), in general, have 
the following relations:
$$
\frac{c_2}{c_3}= - {\rm Tr}[\langle\Phi_e\rangle] , \ \ \
\frac{c_1}{c_3} = \frac{1}{2} \left( {\rm Tr}^2[\langle\Phi_e\rangle] 
-{\rm Tr}[\langle\Phi_e\rangle\langle\Phi_e\rangle] \right) , \ \ \
\frac{c_0}{c_3} = -\det \langle\Phi_e\rangle .
\eqno(A.7)
$$
Therefore, we can predict the ratios $K_e$ and $\kappa_e$ 
as follows:
$$
K_e = 1 -2 \frac{c_1}{c_3} \frac{1}{{\rm Tr}^2[\langle\Phi_e\rangle]} ,
\eqno(A.8)
$$
$$
\kappa_e = -\frac{c_0}{c_3} \frac{1}{{\rm Tr}^3[\langle\Phi\rangle]} ,
\eqno(A.9)
$$
by using the relations (A.7).

Note that the coefficients (A.4) are obtained 
independently of the value of $\varepsilon_{SB}$.
By using a general formula for any $3\times 3$ matrix
$$
L \equiv \frac{{\rm Tr}[\Phi\Phi\Phi]}{{\rm Tr}^3[\Phi]} =
3 \kappa + \frac{3}{2} K - \frac{1}{2} ,
\eqno(A.10)
$$
we can obtain
$$
K_e \equiv \frac{{\rm Tr}[\Phi_e \Phi_e]}{{\rm Tr}^2[\Phi_e]}
= 1+ 2 \frac{c_1}{c_2} \frac{1}{{\rm Tr}[\Phi_e]} 
= 1 -\frac{1}{3} \xi ,
\eqno(A.11)
$$
$$
\kappa_e \equiv \frac{\det \Phi_e}{ {\rm Tr}^3[\Phi_e]}
= \frac{c_0}{c_2} \frac{1}{{\rm Tr}^2[\Phi_e]} 
= \frac{1}{12}(L_e-\xi K_e)
= \frac{1}{4} \left( \kappa_e +\frac{1}{2} K_e 
-\frac{1}{6} -\frac{1}{3}\xi K_e \right) .
\eqno(A.12)
$$
On the other hand, in order to obtain a solution with 
$[\Phi_e]\neq 0$, from the relations (A.4) for 
the coefficients $c_3$ and $c_2$, 
the parameter $\xi$ must be taken as
$$
\xi = 1 .
\eqno(A.13)
$$
Then, we can obtain 
$$
K_e(\Lambda)= \frac{2}{3} ,
\eqno(A.14)
$$
from Eq.(A.11).
Note that the result $K_e(\Lambda)=2/3$ 
has been obtained independently of the parameter 
$\varepsilon_{SB}$.
Again, we would like to emphasize that the result 
$K_e(\mu)=2/3$ is valid only at $\mu=\Lambda$, and
it cannot explain the observed fact $K_e^{pole}= 2/3$.
Therefore, the form of $W_\Phi$ (1.12) is only a toy 
model in the yukawaon approach.

On the other hand,
from Eq.(A.12), we obtain
$$
\kappa_e = \frac{1}{3} \left\{ \left( \frac{1}{2}-
\frac{1}{3}\xi\right) K_e -\frac{1}{6} \right\}
= -\frac{1}{54} .
\eqno(A.15)
$$
The observed value \cite{PDG08}
$$
\kappa^{pole}= (2.0633 \pm 0.0001) \times 10^{-3} ,
\eqno(A.16)
$$
leads to \cite{Xing08}
$$
\kappa_e(\mu)= (2.023 \pm 0.0001) \times 10^{-3}  ,
\eqno(A.17)
$$
at $\mu= 2\times 10^{16}$ GeV for a SUSY model with
$\tan\beta=10$.
Therefore, the toy model (1.12) cannot explain the
observed value of $\kappa$ at all.


\vspace{4mm}

\end{document}